\documentclass[prd,aps,showpacs,floats,floatfix,superscriptaddress,twocolumn,nofootinbib]{revtex4}

\usepackage{psfrag}
\usepackage[dvips]{graphicx}
\usepackage{amsmath}
\usepackage{amsfonts}
\usepackage{color}

\newcommand{\hide}[1]{{}}

\newcommand{\beq}{\begin{equation}}
\newcommand{\eeq}{\end{equation}}
\newcommand{\bea}{\begin{eqnarray}}
\newcommand{\eea}{\end{eqnarray}}
\newcommand{\ba}{\begin{array}}
\newcommand{\ea}{\end{array}}

\newcommand{\nm}{{\rm nm}}
\newcommand{\cm}{{\rm cm}}

\newcommand{\Ein}{E_{\rm in0}}

\newcommand{\ECcira}{E^{\rm C}_{\rm cir1}}
\newcommand{\ECcirb}{E^{\rm C}_{\rm cir2}}
\newcommand{\ECina}{E^{\rm C}_{\rm in1}}
\newcommand{\ECinb}{E^{\rm C}_{\rm in2}}
\newcommand{\ECrea}{E^{\rm C}_{\rm re1}}
\newcommand{\ECreb}{E^{\rm C}_{\rm re2}}

\newcommand{\ECa}{E^{\rm C}_{\rm a}}
\newcommand{\ECar}{E^{\rm C}_{\rm ar}}
\newcommand{\ECs}{E^{\rm C}_{\rm s}}
\newcommand{\ECsr}{E^{\rm C}_{\rm sr}}
\newcommand{\EDout}{E^{\rm S}_{\rm out}}

\newcommand{\EDina}{E^{\rm S}_{\rm in1}}
\newcommand{\EDinb}{E^{\rm S}_{\rm in2}}
\newcommand{\EDrea}{E^{\rm S}_{\rm re1}}
\newcommand{\EDreb}{E^{\rm S}_{\rm re2}}
\newcommand{\EDsiga}{E^{\rm S}_{\rm sig1}}
\newcommand{\EDsigb}{E^{\rm S}_{\rm sig2}}
\newcommand{\EDsigss}{E^{\rm S}_{\rm sigs}}
\newcommand{\EDsigaa}{E^{\rm S}_{\rm siga}}
\newcommand{\EDa}{E^{\rm S}_{\rm a}}

\newcommand{\EDs}{E^{\rm S}_{\rm s}}
\newcommand{\EDsr}{E^{\rm S}_{\rm sr}}
\newcommand{\PCaca}{P^{\rm C}_{\rm ac1}}
\newcommand{\PCacb}{P^{\rm C}_{\rm ac2}}
\newcommand{\PCprc}{P^{\rm C}_{\rm prc}}
\newcommand{\PCsrc}{P^{\rm C}_{\rm src}}
\newcommand{\PCd}{P^{\rm C}_{\rm d}}
\newcommand{\PCmd}{P^{\rm C}_{\rm -d}}
\newcommand{\PDaca}{P^{\rm S}_{\rm ac1}}
\newcommand{\PDacb}{P^{\rm S}_{\rm ac2}}
\newcommand{\PDprc}{P^{\rm S}_{\rm prc}}
\newcommand{\PDsrc}{P^{\rm S}_{\rm src}}
\newcommand{\PDd}{P^{\rm S}_{\rm d}}
\newcommand{\PDmd}{P^{\rm S}_{\rm -d}}
\newcommand{\Mia}{M_{\rm i1}}
\newcommand{\Mib}{M_{\rm i2}}
\newcommand{\Mea}{M_{\rm e1}}
\newcommand{\Meb}{M_{\rm e2}}
\newcommand{\Mp}{M_{\rm p}}
\newcommand{\Ms}{M_{\rm s}}
\newcommand{\MCaca}{M^{\rm C}_{\rm ac1}}
\newcommand{\MCacb}{M^{\rm C}_{\rm ac2}}
\newcommand{\MCC}{M^{\rm C}_{\rm C}}
\newcommand{\MCD}{M^{\rm C}_{\rm D}}
\newcommand{\MDaca}{M^{\rm S}_{\rm ac1}}
\newcommand{\MDacb}{M^{\rm S}_{\rm ac2}}
\newcommand{\MDC}{M^{\rm S}_{\rm C}}
\newcommand{\MDD}{M^{\rm S}_{\rm D}}
\newcommand{\ti}{t_{\rm i}}
\newcommand{\te}{t_{\rm e}}
\newcommand{\tia}{t_{\rm i1}}

\newcommand{\tib}{t_{\rm i2}}

\newcommand{\tp}{t_{\rm p}}
\newcommand{\ts}{t_{\rm s}}
\newcommand{\tbs}{t_{\rm bs}}
\newcommand{\ri}{r_{\rm i}}

\newcommand{\ria}{r_{\rm i1}}
\newcommand{\rea}{r_{\rm e1}}
\newcommand{\rib}{r_{\rm i2}}
\newcommand{\reb}{r_{\rm e2}}
\newcommand{\rp}{r_{\rm p}}
\newcommand{\rs}{r_{\rm s}}
\newcommand{\rbs}{r_{\rm bs}}

\begin{document}

\title{Optimal degeneracy for the signal-recycling cavity in advanced LIGO}

\author{Yi Pan}

\affiliation{Theoretical Astrophysics and Relativity, California
Institute of Technology, Pasadena, CA 91125}

\begin{abstract}
As currently designed, the signal-recycling cavity (SRC) in the Advanced-LIGO interferometer is degenerate. In such a degenerate cavity, the phase fronts of optical fields become badly distorted when the mirror shapes are slightly deformed due to mirror figure error and/or thermal aberration, and this causes significant loss of the signal power and the signal-to-noise ratio (SNR) of a gravitational wave event. Through a numerical modal simulation of the optical fields in a simplified model of an Advanced-LIGO interferometer, We investigate the loss of the SNR and the behavior of both the carrier and signal optical fields, with the SRC at various levels of degeneracy. We show that the loss of the SNR is severe with a degenerate SRC, and a nondegenerate SRC can be used to solve this problem. We identify the optimal level of degeneracy for the SRC, which is achieved when the cavity Gouy phase is between $0.2$ and $1.3$ radians. We also discuss possible alternative designs of the SRC to achieve this optimal degeneracy.

\end{abstract}
\date{July 15, 2006}
\maketitle

\section{Introduction and summary}
\label{sec:intro}

\begin{figure}
\begin{center}
\includegraphics[width=0.5\textwidth]{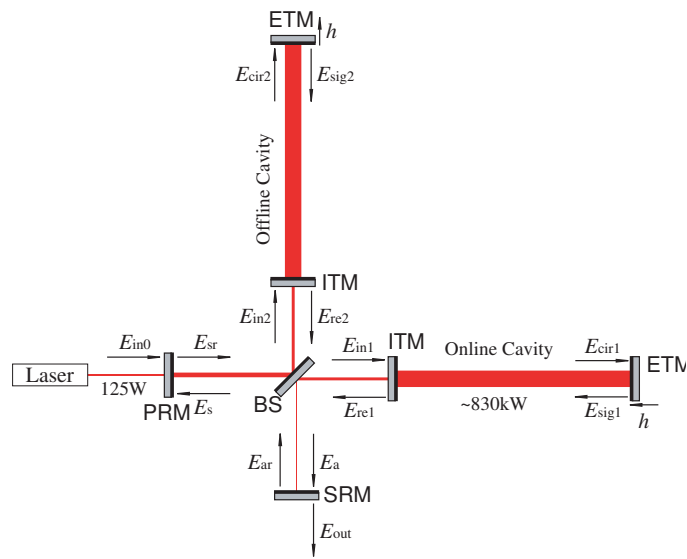}
\caption[Illustration of an Advanced-LIGO interferometer]{Illustration of an Advanced-LIGO interferometer. A signal-recycling mirror is placed at the dark output port of the LIGO interferometer, forming an SRC with the ITMs. In this plot we introduced symbols and abbreviations for the mirrors and the electric fields at various positions that will be used in this chapter. The two ACs are named ``online" and ``offline" for convenience.
\label{fig:ifo}}
\end{center}
\end{figure}

Advanced LIGO \cite{advligo} entails, among other upgrades from initial LIGO \cite{ligo}, introducing a signal-recycling mirror (SRM) at the dark port output of the interferometer  (see Fig.~\ref{fig:ifo}).

The SRM forms the signal-recycling cavity (SRC) with the input test mass (ITM), and the SRC and the arm cavity (AC) form a coupled resonant cavity. The resonant property of this coupled cavity can be controlled by two parameters of the SRM (position and reflectivity) \cite{srcpara,rse}. With different choices of these parameters, the interferometer can operate in either a {\it broadband}, {\it resonant-sideband-extraction} (RSE) configuration \cite{rse, mizuno, heinzel} or a {\it narrowband} configuration. The Advanced-LIGO baseline design adopts the RSE broadband configuration, with the possibility, later, of changing the SRM parameters so as to alter the detector noise spectrum, optimizing its detection of GWs with specific frequency features \cite{advnoise}.

Signal recycling can also circumvent the standard quantum limit (SQL) for free test masses by altering the test-mass dynamics \cite{bc1}.

However, there is a potential problem in the current design of the SRC.

The SRC and the power-recycling cavity (PRC) are both nearly degenerate, and it is well known that a degenerate cavity is not selective for transverse optical modes. As a result, perturbations of the cavity geometry will cause strong mode coupling \cite{degcavity}. Specifically, in initial-LIGO and Advanced-LIGO  interferometers, figure error and thermal aberration of the mirrors (PRM, SRM and ITMs) will cause strong optical mode coupling, which transfers light power from the fundamental TEM00 mode to higher-order modes (HOMs) and reduces the amplitudes of the radio frequency (RF) sidebands (in both the PRC and SRC) and the signal sideband (in the SRC).

This consequence of the high PRC degeneracy is well known: strong mode mixing of the RF sideband has been observed in initial LIGO. This problem was so severe that the interferometer had to be operated with lower circulating power to reduce thermal aberration of the mirrors. Measures had been taken to fix the problem, including introducing a thermal compensation system (TCS) to actively correct the deformations of the mirrors \cite{tcs} and replacing bad optical elements that had unexpectedly high absorption. However, there is a worry about Advanced LIGO, where much higher optical power in the AC will cause even worse thermal aberration on the ITMs that the TCS might not be able to correct. M\"{u}ller and Wise have suggested reducing the PRC degeneracy by moving the mode-matching telescope (MMT) into the recycling cavities to reduce the beam waist size \cite{mmt}, and they are currently working on practical issues in implementing this modified topology in Advanced LIGO \cite{mmtwork}.

The consequence of  SRC degeneracy, by contrast with PRC degeneracy, have not been clearly investigated. Since the GW signal sideband light entering the SRC has different resonance conditions from the carrier light or the RF sideband light, the three must be investigated individually. Specifically, the signal sideband is resonant in the coupled two-cavity system formed by the AC and the SRC, while the carrier and the RF sideband, roughly speaking, are resonant only in the nondegenerate AC and the degenerate PRC respectively. It seems that mode mixing in the signal sideband is at a level somewhere between those of the carrier and the RF sideband.

In Section IV J of Ref.~\cite{mesa}, Thorne estimated the strength of mode mixing in the signal sideband using the approximation that light propagation in the degenerate SRC is well described by geometric optics. (Because a very degenerate cavity accommodates most the optical eigenmodes up to very high orders, ``light rays" with sharp edges are eigenmodes of the cavity as well). Thorne found (Eq. (4.54) of Ref.~\cite{mesa}) that for the signal-to-noise ratio (SNR) to be reduce by less than 1\% due to mode mixing \footnote{Assuming that shot noise is the dominant noise source}, the {\it peak-to-valley} mirror figure error in the central region (region enclosing 95\% of the light power) in the SRC has to be less than 2 nm for the broadband Advanced-LIGO baseline design, and less than 1nm for a narrowband configuration. This is independent of whether a Gaussian beam or Mesa beam \cite{mesa} is used, since a degenerate cavity does not distinguish optical modes. These are severe constraints that are difficult to achieve with current technology.

In this chapter, we investigate this SRC degeneracy problem using a mode-decomposition-based numerical simulation of light propagation, including both the carrier and the signal sidebands, in a simplified model of an Advanced-LIGO interferometer. In our simulations, we focus on the consequences of phase-front distortions of the light in recycling cavities with various levels of degeneracy. We make idealizations and approximations to simplify the analysis of the interferometer, so long as they do not make errors larger than a factor of order two. All assumptions and approximations are discussed in detail in this chapter and we give a full list of them in Appendix~\ref{app:app}. Among the most important ones are the following:

(i) We consider only the five lowest-order HOMs and focus mostly on the two lowest-order modes, i.e., the modes excited by errors in mirror curvature radii.

(ii) We assume that the sizes of mirrors in the interferometers are all large enough for diffraction losses to be negligible.

(iii) We assume that the interferometer noise is dominated by photon shot noise, and we ignore radiation pressure noise. 

(iv) We assume all degrees of freedom of the interferometer (cavity tuning, optical element alignment, etc.) are fixed to their ideal values as if there were no mirror deformation, except for the tuning of the PRC and the AC. The PRC and AC are tuned to maximize the total carrier light power (sum of power in all optical modes) in the AC, for fixed input light power.

We use our simulations to study the loss of the SNR due to phase-front distortions caused by mirror deformations, and reach the following conclusions:

(i) With the degenerate SRC in the current Advanced-LIGO baseline design, if we require the loss of the SNR due to mirror deformations to be smaller than 1\%, then the constraint on mirror deformations is severe. In the broadband Advanced-LIGO design, if the only type of mirror deformation is curvature radius error, then this error must be smaller than 2.5m$\sim$7m on the ITMs, which corresponds to a peak-to-vally figure error smaller than 1nm$\sim$3nm in the central region of the mirrors. In reality, the mirror deformation is formed by the combination of many spatial modes, and when we consider the next lowest order spatial modes, as depicted in Eq.~\eqref{err2} and Fig.~\ref{fig:figerr}, we get a constraint $\sim 4$nm on the SRM. In the narrowband design, the constraint on the ITMs is tightened to $\sim 0.4$nm. These results are consistent with the estimates in Ref.~\cite{mesa}.

(ii) When we change the degeneracy of the SRC while keeping tunings of other cavities the same, we find for Gaussian beams that the loss of the SNR due to mirror deformations is minimized when the one-way Guoy phase inside the SRC is in the range 0.2$\sim$1.3 radians and is chosen to be away from {\it HOM resonant peaks} (see Sec~\ref{sec:result}).

(iii) We find that it is not practical to add a single lens into the SRC so as to focus the beam and reduce the cavity degeneracy to its optimal value, because the beam must be focused so strongly that the beam size on the SRM is of order $10^{-5}$m, and the power density on it exceeds 10GW/m$^2$ (This fact was pointed out long ago by Bochner \cite{bochner}).

(iv) We discuss two alternative designs of the SRC for achieving the optimal degeneracy. The first is the MMT design M\"{u}ller and Wise suggested originally for reducing the degeneracy of the PRC \cite{mmt}. In the MMT design, the beam size is brought down to $\sim10^{-3}$m by two mirrors so as to achieve the optimal degeneracy. The second is a long SRC design, in which the beam size is kept at $\sim$6cm, while the SRC is extended to $\sim$4 kilometers to achieve the optimal degeneracy. The 4km long SRC design has been suggested for various reasons \cite{mizuno,yc}, and in practice it is possible to fit the SRC into the existing LIGO AC beam tubes.

This chapter is organized as follows. In Sec.~\ref{sec:modal}, we give a brief overview of the mode decomposition formalism and Hermite-Gaussian modes \cite{laser}, and interpret the cavity degeneracy from a modal-space point of view. In Sec.~\ref{sec:ifomodel}, we describe the Advanced-LIGO interferometer model that is used in our simulations. In Sec.~\ref{sec:result}, we summarize the numerical results that come out of our simulations, including the constraints on mirror figure error and thermal aberration, and the optimal SRC degeneracy. In Sec.~\ref{sec:designs}, we discuss the various designs for achieving the optimal SRC degeneracy. In Sec.~\ref{sec:conc}, we summarize our conclusions.

\section{Mode decomposition formalism}
\label{sec:modal}

\subsection{Modal decomposition in general}
\label{modal}
The mode decomposition formalism for calculating optical fields in a perturbed interferometer is discussed in detail by Hefetz {\it et al} in Sec. 2 of Ref.\cite{principle}. We will review the general idea briefly in this section. 

One can generally expand the electromagnetic (EM) field of a light beam as a superposition of orthonormal optical modes:
\beq
E(x,y,z)=\sum_na_nU_n(x,y,z) \,.
\eeq
Though the basis modes $U_n(x,y,z)$ are arbitrary in principle, it is preferable to use the eigenmodes of the cavities of an ideal interferometer, e.g., (i) Hermite-Gaussian modes, which are eigenmodes of the cavities in the Advanced-LIGO baseline design, formed by spherical mirrors (assuming infinite mirror size); or (ii) Mesa-beam modes \cite{mesa}, which are eigenmodes suggested for Advanced LIGO to reduce thermal noise. The complex vector space formed by $U_n(x,y,z)$ is called the {\it modal space}, and the EM field in modal space is represented by a complex vector $a_n$. (In Sec.~\ref{sec:ifomodel} and Appendix~\ref{app:fields} we keep using $E$ to denote this vector.) In the modal space, the optical fields of a perturbed interferometer can be calculated from the unperturbed fields using linear algebra only, without numerically solving the wave equation.

The propagation of the optical field can be described by matrices in this modal space. In Cartesian coordinates where the $z$-axis is along the optical axis and the $x$- and $y$-axis are transverse, an operator $M(x,y,z_2,z_1)$ transforms the EM field at position $z_1$ to the field at position $z_2$:
\beq
E(x_2,y_2,z_2)=M(x_2,y_2;x_1,y_1;z_2,z_1)\otimes E(x_1,y_1,z_1) \,.
\eeq
The representation of $M$ in the modal space is given by
\begin{multline}
M_{mn}(z_2,z_1)=\iiiint_{-\infty}^{\infty}U^*_m(x_2,y_2,z_2) \\
\times M(x_2,y_2;x_1,y_1;z_2,z_1)U_n(x_1,y_1,z_1)dx_1dy_1dx_2dy_2 \,.
\end{multline}

It is convenient to separate these operators into propagation operators in free space and interaction operators describing how the EM fields transform when interacting with optical elements. The free-space propagator is given by:
\beq\label{prop}
P_{mn}(z_1,z_2)=\delta_{mn}e^{-ik(z_2-z_1)}e^{i\eta_n}\,,
\eeq
where $k$ is the wave number and $\eta_n$ is the diffraction phase associated with the $n$th optical mode (i.e., the extra phase accumulated during propagation besides $k(z_2-z_1)$, due to diffraction effects, e.g., the Guoy phase of Gaussian beam).

To write out the interaction operator for an optical element, we choose, near the element's surfaces, reference surfaces that match the phase fronts of the unperturbed eigenmodes. The operator can then be written in the general form
\beq\label{inter}
M_{mn}=\langle m|M(x,y)|n\rangle=\langle m|e^{-ikZ(x,y)}|n\rangle\,,
\eeq
where $Z(x,y)$ is the optical path light travels from the element's entrance reference surface to the element's exit reference surface. For ideal optical elements in our model, $Z(x,y)={\rm constant}$, i.e., the mirrors or lens exactly match the optical modes and there is no coupling between optical modes when the light beam interacts with the optical elements. For perturbed elements, e.g., slightly deformed mirrors due to figure error and/or thermal aberration, $Z(x,y)$ is not constant and the optical modes couple to each other. $Z(x,y)$ contains contributions from both the figure error and the change of refraction index in the material, and is referred to as the {\it distortion function}; it can be complex, when used to describe lossy optical elements. This interaction operator must be accompanied by the optical element's scalar reflectivity and transmissivity coefficients to give the true transform of the fields.

In writing the interaction operators, we adopted the short-distance approximation \cite{VHB}, where propagation inside optical elements between reference planes is approximated by a simple non-uniform phase factor $kZ(x,y)$. The spatially variable phase error caused by this approximation, in addition to a factor of the order unity determined by the geometry of the unperturbed cavity, is derived in Sec. 2 of Ref.~\cite{VHB}; its magnitude is
\beq
\Delta\Phi\sim\frac{1}{4\pi}\frac{\lambda}{L}\,,
\eeq
where $\lambda$ is the wavelength of the light and $L$ is the length of the cavity. In initial LIGO or Advanced LIGO, $L$ is at least $\simeq10$m, and the phase distortion error is thus smaller than $10^{-8}$. Since a 1nm mirror figure error gives a phase distortion of $6\times 10^{-4}$, $\Delta\Phi$ is always negligible.

\subsection{Hermite-Gaussian modes}
\label{gaussian}

In this section, we review briefly the Hermite-Gaussian modes (see Chapters 16 and 17 of Ref.~\cite{laser}) that are used as basis modes in our simulation.

A Hermite-Gaussian mode of beam waist size $w_0$ is given by
\bea\label{hgmode}
U_n(x,z)&=&\left(\frac{2}{\pi}\right)^{1/4}\left(\frac{1}{2^nn!w(z)}\right)^{1/2}H_n\left(\frac{\sqrt{2}x}{w(z)}\right) \nonumber\\
&&\times{\rm exp}\left(-x^2\left(\frac{1}{w(z)^2}+\frac{ik}{2R(z)}\right)\right) \nonumber\\
&&\times{\rm exp}\left(i\left(n+\frac{1}{2}\right)\eta(z)\right) \,,
\eea
where $H_n(x)$ is the Hermite polynomial and $R(z)$, $w(z)$, and $\eta(z)$ are the curvature radius of the phase front, the beam spot size and the Guoy phase, respectively, given in terms of the Rayleigh length $z_0=\pi w_0^2/\lambda$ by
\bea
R(z)=z+\frac{z_0^2}{z}\,, & \qquad & w(z)=w_0\sqrt{1+\frac{z^2}{z_0^2}} \,, \nonumber\\
\mbox{and} && \eta(z)=\tan^{-1}\left(\frac{z}{z_0}\right) \,.
\eea
These Hermite-Gaussian modes are exact solutions to the paraxial wave equation in one dimension, and they form a complete orthogonal basis in the solution space. We use Hermite-Gaussian modes as basis modes in the modal space
\beq
E(x,y,z)=\sum_{mn}a_{mn}U_m(x,z)U_n(y,z){\rm exp}(-ikz) \,,
\eeq
where each transverse mode is labeled by two integers ($m,n$) corresponding to the directions $x$ and $y$. In this chapter, we consider only the five lowest-order symmetric HOMs, i.e., modes with even $m$ and $n$, and $m+n\le4$. We omit the two $m+n=1$ modes because they can be corrected by the tilt control system. We omit modes with $m+n>4$ to reduce computational cost, and because modes with higher orders are suppressed by stronger diffraction loss, their coupling to the fundamental mode is weaker when mirror deformations are smooth. Moreover, mode-coupling behaviors due to mirror deformations are qualitatively the same for all modes, and there is no need to include more modes with order $m+n>4$ for an order of magnitude estimation. We also focus our attention on the second order ($m+n=2$) modes since they are most likely to be the dominant deformations present in Advanced LIGO due to mirror figure error or thermal lensing, and thus are the lowest order perturbations of Hermite-Gaussian modes after mirror tilts have been suppressed by control systems.

The propagation operators for Hermite-Gaussian modes are given by Eq.~\eqref{prop}, with the diffraction phases replaced by the Gouy phases of the ($m,n$) Hermite-Gaussian modes: $(m+n+1)\eta(z)$. The interaction operators defined in Eq.~\eqref{inter} are derived analytically for Hermite-Gaussian modes in Section 2 of Ref.~\cite{principle}, assuming that the mirror radius is much larger than the beam size. For the mirror deformations considered in this chapter, this approximation produces a few percent errors in the coupling strengths between optical modes (measured by components of the interaction operators). In fact, these errors are less than $10^{-3}$ for the coupling between the fundamental mode and the five HOMs we considered, which is the leading order effect we would like to investigate in this chapter. Only for coupling between HOMs, which is a higher-order effect in changing the SNR, does the error caused by the finite size of mirrors become as large as $1\sim 10$\%. Also, because Gaussian modes are eigenmodes of cavities formed by spherical mirrors with infinite size, it is in fact self-consistent that we treat the mirrors as large when using Gaussian modes as eigenmodes. As long as we use eigenmodes consistent with mirror size, even if we use infinite-size mirrors, our estimates are valid up to fractional errors of order the difference between the Hermite-Gaussian modes and the true eigenmodes of the cavity with finite-size mirrors. Thus, we approximate the mirrors as having infinite sizes and use Gaussian modes throughout the chapter.

Gaussian beams have spherical phase fronts, and are thus eigenmodes of cavities formed by spherical mirrors. There are simple relations between the geometry of the cavity (measured by, e.g., cavity {\it g-factor}, and mirror radii of curvature) and the physical properties of the Gaussian eigenmodes (e.g., Guoy phase, beam waist size, waist position) that are available in Chapter 19 of Ref.~\cite{laser}. 

A very useful formula relating a cavity's one-way Guoy phase $\eta$ (the Guoy phase of the cavity's fundamental eigenmode) to the cavity g-factor is:
\beq
\eta=\arccos\sqrt{g} \,.
\eeq
The cavity g-factor is the product of the two mirror g-factors. If the curvature radii of the two mirrors are $R_1$ and $R_2$, and the length of the cavity is $L$, then the mirror and cavity g-factors are defined as
\beq
g_i\equiv 1-\frac{L}{R_i} \quad (i=1,2) \quad\mbox{and}\quad g\equiv g_1g_2 \,.
\eeq
The cavity is stable (i.e., the cavity geometry supports Hermite-Gaussian modes as eigenmodes) if and only if $0\le g\le1$.

From Eq.~\eqref{prop}, we see that it is the Guoy phase that distinguishes optical modes (modes with different $m+n$ for Hermite-Gaussian modes). If the one-way cavity Gouy phases are very close to 0 or $\pi$, the round trip phase shifts in the cavity are almost the same for all modes, so if one of the optical modes is tuned to be resonant, so are the others. Such a cavity is thus called degenerate. In terms of g-factor, a cavity is degenerate if $g$ is very close to $0$ or $1$.

To conclude this section, we use the baseline design of the AC and the PRC in Advanced LIGO as examples to demonstrate quantitatively the degeneracy level of cavities. The baseline curvature radii of the PR mirror and the test masses are \cite{mmt}
\bea\label{mirrorR}
&&R_{\rm ETM}=R_{\rm ITM}=2076.4{\rm m} \,, \qquad  R_{\rm ITM2}=-1186.4{\rm m} \,,\nonumber\\
&&R_{\rm PR}=1194.7{\rm m} \,,
\eea
where $R_{\rm ITM}$ and $R_{\rm ITM2}$ are curvature radii of the ITM seen from inside the AC and the PRC, respectively, and the ITM is convex as seen from the PRC. The cavity lengths are $d_{\rm AC}=4000$m and $d_{\rm PRC}=8.34$m, and the Rayleigh lengths and Guoy phases are
\bea
z_{0\,\rm AC}= &390.9{\rm m} \,,\qquad \eta_{\rm AC}=&0.39 \,, \\
z_{0\,\rm PRC}= &82.1{\rm m} \,,\qquad \eta_{\rm PRC}=&4.9\times10^{-4} \,.
\eea
In the AC, the Rayleigh length is clearly much shorter than the typical distance 4km that carrier light travels in the cavity, i.e., the light propagation is in the strong-diffraction zone, which indicates a nondegenerate cavity. More rigorously, the Guoy phase, corresponding to a frequency shift of
\beq
\Delta\nu=\frac{c}{2\pi d_{\rm AC}}\eta_{\rm AC}=4.6{\rm kHz} \,,
\eeq
is much larger than the bandwidth of the AC ($\sim 15$Hz). This means, the Guoy phase breaks the degeneracy between the Gaussian modes with different orders (different $m+n$), i.e., when the cavity is tuned to have the fundamental mode in resonance, nearly all other HOMs are suppressed. Of course there are always HOMs with round trip Guoy phases (mode $2\pi$) close to that of the fundamental mode by coincidence, but these HOMs that resonate with the fundamental mode generally have very high orders (except for very bad choices of the Guoy phase) and are thus strongly suppressed by diffraction losses.

In the PRC, the Rayleigh length is longer than the length of the cavity, but shorter than the typical distance RF sideband light travels inside the cavity after we take count of the number of round trips ($\sim$ 50). Therefore, the RF sideband propagation in the PRC is still in its strong diffraction zone. However, the frequency shift of $\Delta\nu=2.8$kHz is much smaller than the bandwidth of the PRC ($\simeq 100$kHz), and consequently the Guoy phases are close to 0 and $\pi$ (mode $2\pi$). The PRC, therefore, although not degenerate to the extreme level that geometric optics becomes valid, accommodates tens of low-order HOMs together with the fundamental mode, and is thus highly degenerate.

Finally, through the example above, we can see that for Hermite-Gaussian modes, reducing the beam waist size $w_0$ (i.e., reducing Rayleigh length $z_0$) and/or increasing the cavity length will reduce the degeneracy of the cavity.

\section{Advanced-LIGO interferometer modeling}
\label{sec:ifomodel}

In this section, we describe our simplified model of an Advanced-LIGO interferometer, and the way our simulations work.

In our simulations, we study an Advanced-LIGO interferometer in equilibrium with static optical fields. We use the standard Advanced-LIGO optical topology displayed in Fig.~\ref{fig:ifo}, and the input light is a pure (0,0) Hermite-Gaussian mode coming in from the PR mirror. We consider both broadband and narrowband interferometer designs. The interferometer parameters for the broadband detector are chosen as their values for the Advanced-LIGO baseline design. The parameters in the two designs are listed below \cite{mmt,mesa,advligoref}, where we begin using the following subscripts to denote different mirrors and cavities throughout the whole chapter (see Appendix ~\ref{app:symbol} for a list of symbols and subscripts):
``bs," ``i," ``e," ``p," and ``s" stand for the beam splitter (BS), ITM, ETM, PRM, and SRM; ``ac," ``prc," and ``src" stand for AC, PRC, and SRC.

(i) {\it Cavity macroscopic length}: The ACs both have $L=4000$m; the lengths between the PRM and the two ITMs are denoted $l_1$ and $l_2$ and referred to as Michelson lengths; the lengths between the SRM and the two ITMs are denoted $l_3$ and $l_4$. It is convenient to define common and differential lengths:
\beq\label{com}
l_{\rm prc}\equiv\frac{l_1+l_2}{2}=8.34{\rm m} \,,\qquad l_{\rm src}\equiv\frac{l_3+l_4}{2}=8.327{\rm m} \,,
\eeq
and
\beq
d\equiv\frac{l_1-l_2}{2}=\frac{l_3-l_4}{2} \,.
\eeq\label{dif}
In our model, there is no macroscopic asymmetry between the two Michelson arms. Therefore only a microscopic tuning value is assigned to $d$ in the next paragraph. 

(ii) {\it Cavity microscopic tuning}: The carrier light gets the following phase shifts during a single trip in the AC, PRC, and SRC:
\beq\label{tuning}
\phi_{\rm ac}=\phi_{\rm pc}=0 \,,\qquad \phi^{\rm B}_{\rm src}=0.06 \,,\qquad \phi^{\rm N}_{\rm src}=\pi-1.556 \,,
\eeq
where the superscripts on the SRC phasing denote broadband (B) and narrowband (N). An asymmetry in the Micheleson arm lengths is introduced because, among other reasons, we choose the homodyne readout scheme where a tiny amount of the carrier light power goes toward the dark port and beats with the resonant signal sideband to give the detector output. For this purpose solely, the asymmetry is specified at the microscopic level as the phase difference the carrier accumulates in the two Micheleson arms: $\displaystyle\Delta\phi=\omega_0 d/c=0.01$, so that about 1W of carrier power goes into the SRC.

(iii) {\it Mirror power transmissivity}:
\bea\label{trans}
\ti^2=&0.5\% \,,\qquad \te^2=&76\,{\rm ppm} \,,\qquad \tp^2=5.9\% \,,\qquad \nonumber\\
{\ts^{\rm B}}^2=&7\% \,,\qquad\;\; {\ts^{\rm N}}^2=&0.3\% \,.
\eea
We assume lossless mirrors throughout our simulation, so the amplitude reflectivity and transmissivity are completely determined.

(iv) {\it Mirror curvature radii}: These are given in Eq.~\eqref{mirrorR} except for the SRM. In this chapter, we will change the SRC degeneracy through the value of its Guoy phase, and assume that the geometry of the SRM always matches our choice of degeneracy. The corresponding SRM curvature radius is relevant to nothing but the SRC Guoy phase, so we do not specify it explicitly

The differences between the broadband and the narrowband designs are all in the choice of the SRM transmissivity and the SRC tuning, as we mentioned in Sec.~\ref{sec:intro}. The complex optical-sideband resonant frequency in the coupled SRC and AC two-cavity system is given by Eq. (13) of Ref.~\cite{scaling}:
\beq\label{omegas}
\tilde{\omega}=\frac{ic}{2L}\log\frac{\ri+\rp e^{2i\phi_{\rm src}}}{1+\ri\rp e^{2i\phi_{\rm src}}}\equiv -\lambda_g-i\epsilon \,,
\eeq
where $\lambda_g$ and $\epsilon$ are positive and are the resonant frequency and the decay time. With our choice of signal-recycling parameters above, we get resonant sideband frequencies $\lambda_g/2\pi=228$Hz and $\lambda_g/2\pi=1005$Hz. The actual resonant frequency is $\omega_0-\lambda_g$, i.e., the down-converted signal sideband. Depending on the sign of the SRC detuning, the interferometer response is of interest for only one of the two sidebands.

The PRM transmissivity and the PRC tuning are chosen such that the PRM is impedance matched to the AC and the total carrier power (summed over all optical modes) in the AC is maximized for fixed input light power.

The only interferometer control we do in simulations when mirrors are slightly deformed is to optimize the total carrier light power in the AC by adjusting the tuning of the PRC and the AC. To optimize the carrier light power in the AC, instead of modeling the control signal, we look directly at the power at each equilibrium state, i.e., we do a static pseudo-control and do not model the dynamical response of the interferometer during the control process. With our choice of interferometer parameters given above, and a 125W input light power, the carrier light power in an ideal interferometer is $\simeq 825$kW, the power recycling factor is $\simeq 18$, and the carrier light going toward the dark port is $\simeq 1$W.

Since we are interested in the reduction of the signal sideband power due to mirror deformations, we consider in our simulations only the carrier light and the down-converted signal sideband, and omit the RF sideband and other sidebands used for control purposes, as well as the up-converted signal sideband, since the interferometer is tuned to be sensitive only to the down-converted signal sideband with the best sensitivity (assuming white noise) at frequency $f_0-\lambda_g/2\pi$. The distortion of the phase fronts also affect the control sidebands, but we are not concerned with several percent loss of the SNR of the control signals. However, there is a problem associated with tilt control signals, whose SNRs are proportional to the amplitude of the (1,0) and (0,1) Hermite-Gaussian modes excited by mirror tilts. If we choose to suppress HOMs for the signal sideband by using nondegenerate recycling cavities, the SNR of the tilt control signals entering the recycling cavities will also be strongly reduced. Correspondingly, choosing the degeneracy level of the SRC will entail compromises between inputs on the signal and control sidebands. This problem is not considered in this chapter, and is left for future investigation.

After setting up our model of an ideal interferometer, we introduce perturbations to its mirrors. We do not model thermal lensing of the mirrors. We assume that all tilts and misalignments of the mirrors are corrected by the control system. Because of the high computational cost associated with our mode decomposition method, we limit our figure errors to simple profiles so that they generate, at the leading order, coupling between only a few optical modes ($\le 6$). One focus of our study is mirror curvature radius errors, which is the most interesting type of deformation, since it can be generated effectively by thermal lensing of the ITMs. We assume that the beamsplitter is perfect, because deformations of the beamsplitter introduce complicated phase-front distortions which have no qualitative difference from those introduced by other mirrors.

At the interferometer output, we assume that there is a mode cleaner that filters out all HOMs in the carrier and signal sidebands, so that the shot noise is proportional to the square root of the output carrier power in the fundamental mode [i.e., (0,0) Gaussian mode]:
\beq
N_{\rm shot}\propto\sqrt{I^C_{00}} \,.
\eeq
Here the superscript ``C" stands for carrier, and the subscript labels the mode.
The signal power comes from beating the signal sideband against the carrier, both taking only the fundamental mode, so
\beq
S\propto\sqrt{I^C_{00}}\sqrt{I^S_{00}} \,,
\eeq
where the superscript ``S" stands for signal sideband. Assuming shot noise dominates, we have
\beq\label{snrtosig}
{\rm SNR}\propto\sqrt{I^S_{00}} \,,
\eeq
i.e., the SNR is directly proportional to the signal sideband amplitude in the fundamental mode. 

When we take into account radiation-pressure noise, the change of the SNR becomes more delicate. However, since the radiation-pressure noise is determined by the total light power on the test masses, it is presumably less sensitive to the mode structure of the light. When radiation-pressure noise is important, although the loss of the SNR is not given by Eq.~\eqref{snrtosig}, that equation is still a rough measure of the loss of the SNR. This argument applies to all other non-optical noise sources. Radiation-pressure noise is thus omitted in our simplified model, and will be studied in more sophisticated future simulations \cite{fft}.

In the interferometer model described above, we calculate the signal sideband in two steps. In the first step, we propagate the input carrier light (Nd:YAG laser) with frequency $f_0=2.82\times 10^{14}$Hz through the interferometer to build up the static carrier-light field. In the second step, we assume a sinusoidal gravitational wave of frequency $f_g$ propagating perpendicular to the detector plane with only ``+" polarization, i.e., effectively, it differentially shakes the ETMs sinusoidally with frequency $f_g$. To leading order in the GW strain, two signal sidebands of frequencies $f_0\pm f_g$ are generated at the ETMs with exactly the same mode structures as the carrier field there. We propagate the down-converted sideband through the interferometer to build up the static signal sideband field. Repeating this second step with various GW frequencies, we map out the frequency response of the detector. Repeating both steps with various SRC geometries and levels of degeneracy, we can study the effect of the recycling-cavity degeneracy on the influence of mirror deformations on signal response and the interferometer's noise spectrum.

The EM field in the interferometer (a system of coupled optical cavities) can be written formally as
\beq\label{Eeqn}
E=E_{\rm pump}+P_{\rm r.t.}E \,.
\eeq
Here $E_{\rm pump}$ is the pumping field that contributes directly to the $E$ field, and $P_{\rm r.t.}$ is the round-trip propagator, which consists of free propagation operators and interaction operators that describe the propagation of $E$ through the interferometer and back to itself. The specific forms of $E_{\rm pump}$ and $P_{\rm r.t.}$ depend on location in the interferometer. For example, the field $E_{\rm cir1}$ in Fig.~\ref{fig:ifo}, i.e., the circulating field in the online AC at the ITM going toward the ETM, can be written as
\beq
E_{\rm cir1}=t_{\rm i}T_{\rm i1}E_{\rm in1}+r_{\rm i}r_{\rm e}M_{\rm i1}P_{\rm ac1}M_{\rm e1}P_{\rm ac1}E_{\rm cir1} \,,
\eeq
where, e.g., $M_{\rm i1}$ and $T_{\rm i1}$ are reflection and transmission operators of the ITM in the online cavity, and we use subscripts ``1" and ``2" to denote the online and offline cavities.
We can write out a set of coupled equations in the form of Eq.~\eqref{Eeqn} for all fields labeled in Fig.~\ref{fig:ifo}, and solve them numerically by iteration, as has been done in the FFT simulation code for optical fields in LIGO \cite{bochner}. In principle, we can also solve for each field in terms of the input field $\Ein$ by directly taking the inverse of all operators of the form $(I-P_{\rm r.t.})$ ($I$ is the identity matirx). In the FFT code, hundreds of modes are included, and it is computationally difficult to take the inverse of all the large matrices, which are sometimes nearly singular. In our simulation however, as we consider only six modes, we find it more efficient to directly invert the matrices instead of iterating the fields.

In Appendix~\ref{app:fields}, we write out the field coupling equations explicitly and solve for all the carrier and signal sideband fields.

\section{Mirror figure error and optimal degeneracy}
\label{sec:result}

In this section, based on results of our simulations of the simplified Advanced-LIGO model set up in Sec.~\ref{sec:ifomodel}, we try to answer the question {\it how much loss of the SNR is caused by mirror deformations with various spatial modes and magnitudes, and how does this loss of the SNR depend on the degeneracy of the SR cavity?}

For mirror deformations, we consider mostly mirror curvature-radius errors. At leading order, we need only consider the (2,0) and (0,2) Hermite-Gaussian modes excited by curvature radius error.

We also consider a case in which, besides the curvature radius error, there is deformation of the SRM with higher-order spatial modes. We are interested in this case since, if the SRC is nondegenerate, the optical eigenmodes of the SRC and the coupled SRC-AC cavities under the mirror deformation are different from those of the coupled PRC-AC cavity, and there might be a substantial loss of the SNR due to the mode mismatch between the carrier light (mostly in the PRC-AC cavity) and the signal sideband light (mostly in the SRC-AC cavity). Thus we would especially like to see how the loss of the SNR depends on the degeneracy of the SR cavity in this case.

For signal sidebands, we consider mostly the resonant signal sideband with frequency $f_0-\lambda_g/2\pi$ given by Eq.~\eqref{omegas}, and only at the end of this section do we consider signal sidebands with frequencies varying from $50$Hz to $1000$Hz.

\subsection{Curvature radius error on the ITMs: Broadband configuration}
\label{itm2broad}

\begin{figure}
\begin{center}
\includegraphics[width=0.5\textwidth]{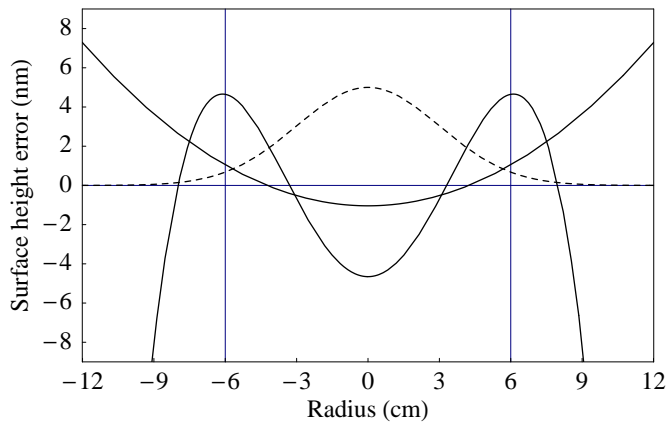}
\caption[Two examples of mirror figure error]{Mirror figure errors given in Eqs.~\eqref{err1} and \eqref{err2}. We show as solid curves the error $\Delta z$ in units of nanometers along the $x$-axis [i.e., $\Delta z(x,0)$]. The dashed curve is the power profile of the fundamental Gaussian mode plotted in arbitrary units. The beam size at the mirror is $w=6$cm (c.f. Eq.~\eqref{hgmode}), so the variance of the power profile is $\sigma=3$cm. The peak-to-valley figure errors inside the $2\sigma$ (6cm) region are about $2.0$nm and $9.3$nm for the two cases.
\label{fig:figerr}}
\end{center}
\end{figure}

First, we consider a broadband interferometer with curvature radius errors on the ITMs that simulate the thermal lensing effect, and we assume all other mirrors are perfect. The curvature radii of the ITMs ($R_{\rm ITM}=2076.4$m) are changed by $\Delta R_{\rm ITM}=5$m, either commonly or differentially. This curvature radius error corresponds to the following mirror figure error:
\beq\label{err1}
\Delta z(x,y)=1.04{\rm nm}\left[-1+2\left(\frac{x}{6{\rm cm}}\right)^2+2\left(\frac{y}{6{\rm cm}}\right)^2\right] \,,
\eeq
where $\Delta z(x,y)$ is the mirror surface height error, and $\Delta z(x,0)$ is show in Fig.~\ref{fig:figerr}.

\begin{figure*}
\begin{center}
\includegraphics[width=1\textwidth]{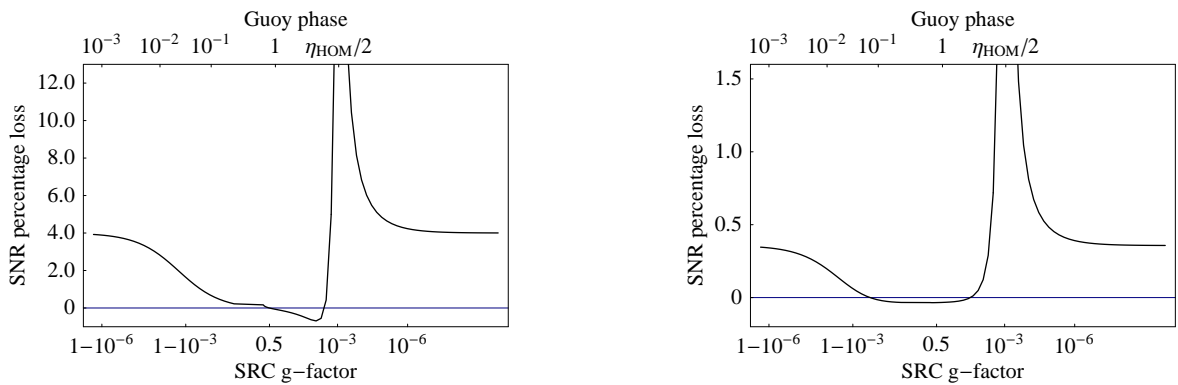}
\caption[Loss of the SNR in Advanced-LIGO interferometers due to mirror curvature-radius errors on the ITMs]{Loss of the SNR in Advanced-LIGO interferometers due to mirror curvature radius error on the ITMs, at the resonant signal sideband frequency $f_0-\lambda/2\pi$, as a function of the SRC degeneracy level. The curvature radius of the ITMs is $R_{\rm ITM}=2076.4$m and we consider an error $\Delta R_{\rm ITM}=5$m, which is equivalent to the mirror surface height error given in Eq.~\eqref{err1}. We consider the two cases where the curvature errors on the two ITMs are common (i.e., curvature radii are $R_{\rm ITM}-\Delta R_{\rm ITM}$) and differential (i.e., curvature radii are $R_{\rm ITM}\pm\Delta R_{\rm ITM}$), and show the loss of the SNR in the left and right panels, respectively. The horizontal axis is the level of degeneracy, measured by two quantities: the g-factor of the SRC and the one-way Guoy phase in the SRC. The horizontal position of the left-most point of the curve corresponds to the degeneracy level of the SRC in the current Advanced-LIGO baseline design.
\label{fig:itm2sig}}
\end{center}
\end{figure*}

We show the loss of the SNR in Fig.~\ref{fig:itm2sig}. With changes of the SRC degeneracy, characterized by the Guoy phase and the SRC g-factor, we see significant change in the loss of the SNR. In the baseline degenerate design, for common and differential perturbations we lose 4\% and 0.4\% of the SNR. Note that at leading order the SNR loss is proportional to the square of the size of the error, so we have, for instance, 4 times the above SNR losses when $\Delta R_{\rm ITM}=10$m instead of 5m. The bigger of these SNR losses is consistent with the estimate based on geometric optics approximation in Section IV J of Ref.~\cite{mesa}.

Figure \ref{fig:itm2sig} shows that, when the degeneracy is reduced, the SNR loss drops by more than two orders of magnitude, making the curvature error harmless. The most striking features in the plots are the peaks corresponding to huge SNR loss at some nondegenerate SRC configuration. This happens when the Guoy phase of the HOMs (in this case, (2,0) and (0,2) Hermite-Gaussian modes) cancels the SRC detuning ($\phi^{\rm B}_{\rm src}=0.06$ for broadband design), so that the HOMs of both the carrier and the signal light are resonant in the SRC while the fundamental modes are detuned. This is clearly a bad choice of SRC degeneracy. We refer to it as the HOM resonant peak. When more HOMs are coupled into the interferometer by perturbations, we should avoid all such cavity configurations in which some HOM (of an order not so high that it suffers strong diffraction loss) has a Guoy phase $\eta_{\rm HOM}$ that nearly cancels the SRC tuning phase.

\begin{figure*}
\begin{center}
\includegraphics[width=1\textwidth]{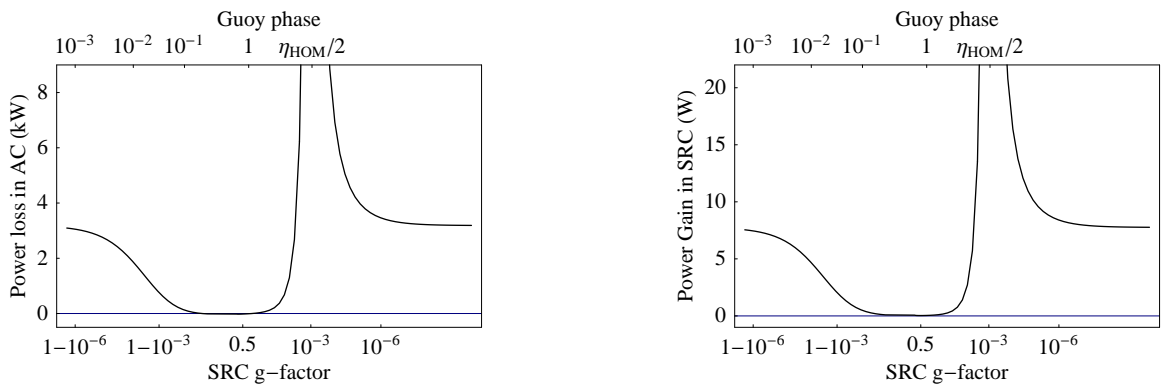}
\caption[The change of the carrier light power in the AC and SRC in Advanced-LIGO interferometers due to differential mirror curvature errors on the ITMs]{Change of the carrier light power in the AC and SRC in Advanced-LIGO interferometers due to differential mirror curvature radius error on the ITMs, as a function of SRC degeneracy. We consider the same differential curvature error as in Fig.~\ref{fig:itm2sig}. In the left panel, we show the loss of carrier power (all optical modes) in the AC in units of kW (the ideal AC carrier power is about 825kW); in the right panel we show the increase of the carrier power (all optical modes) in the SRC in units of W (the ideal SRC carrier power is about 1W). The horizontal axis is the level of degeneracy, measured by two quantities: the g-factor of the SRC and the one-way Guoy phase in the SRC. The horizontal position of the left-most point of the curve corresponds to the degeneracy level of the SRC in the baseline Advanced-LIGO design.
\label{fig:itm2car}}
\end{center}
\end{figure*}

Figure \ref{fig:itm2car} shows the effect of the SRC degeneracy on the loss of carrier light due to differential curvature errors on the ITMs. In addition to the HOM resonant peaks discussed above, there are other noticeable features in the carrier light. In the AC, curvature errors reduce the carrier-power build-up in ways described below. When the SRC becomes nondegenerate, for common and differential curvature errors on the ITMs, the carrier and signal sideband light behaves very differently (compare Figs.~\ref{fig:itm2sig} and \ref{fig:itm2car}).

For common curvature errors, the HOMs are coupled into the symmetric port of the interferometer, which is accepted by the degenerate PRC but are anti-resonant in it, so they get reflected back into the AC. Since the HOMs do not enter the SRC, its degeneracy causes no difference. In this case the carrier light power is controlled by the AC and is hardly affected by the mirror deformations, thus we do not show it. The SNR loss in this case is due directly to the coupling of signal-light HOMs to the dark port.

For differential curvature error, by contrast, the HOMs are coupled to the anti-symmetric port of the interferometer, and when the SRC is degenerate, because the carrier light is not anti-resonant in it, it behaves like ``resonant carrier extraction," and sucks carrier light power out of the AC thus reducing its power build-up. When the SRC is nondegenerate, HOMs are rejected by both the SRC and the AC, and the carrier power builds up as usual in the AC in almost entirely its fundamental mode, without losing any significant power. The SNR loss in this case is due to the loss of carrier power in the AC, since the signal sideband HOMs are coupled to the symmetric port and rejected by the PRC, which is not properly detuned for the signal sideband frequency.

In conclusion, common and differential errors in the ACs reduce the SNR in different ways, and the differential errors reduce the carrier build-up significantly. Finally, as indicated above by ``resonant carrier extraction," the differential curvature error sends a huge amount of carrier power in HOMs toward the dark port ($7.5$W in HOMs and $0.3$W in fundamental modes inside the SRC for $\Delta R_{\rm ITM}=5$m), and thus sends reference light that is mostly in HOMs toward the photon detector. This HOM light must be cleaned out by an output mode cleaner.

\begin{figure}
\begin{center}
\includegraphics[width=0.5\textwidth]{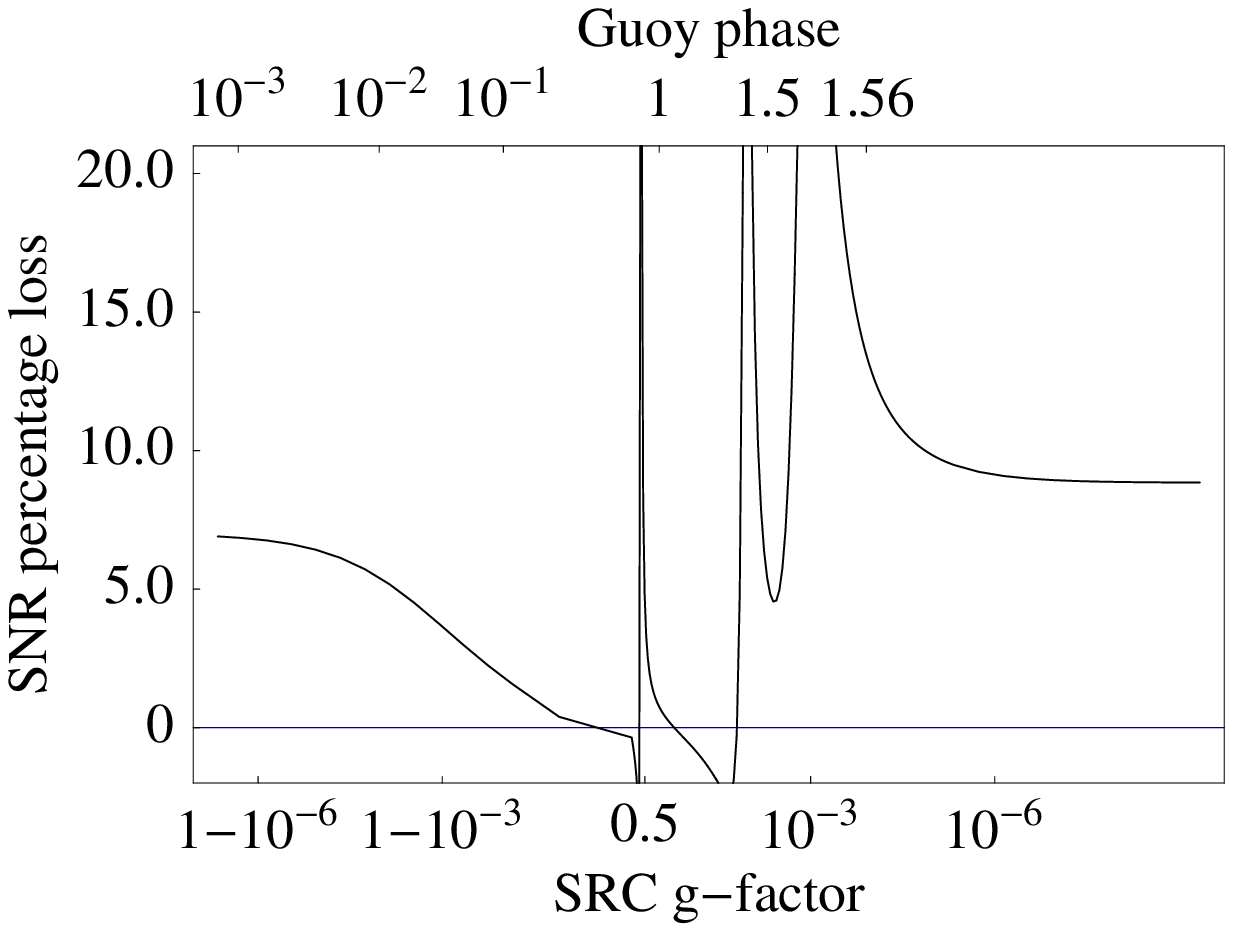}
\caption[Loss of the SNR in Advanced-LIGO interferometers due to mirror curvature radius error on the ITMs and the SRM, and higher-order mode deformation on the SRM]{Loss of the SNR in Advanced-LIGO interferometers due to mirror curvature radius error on the ITMs and the SRM, and some higher-order mode deformation on the SRM, at the resonant signal sideband frequency $f_0-\lambda/2\pi$, as a function of SRC degeneracy. We consider a common curvature radius error on the ITMs with $\Delta R_{\rm ITM}=1$m, i.e., one-fifth the error considered in Fig.~\ref{fig:itm2sig} and \ref{fig:itm2car} or equivalently one-fifth the mirror surface height error given by Eq.~\eqref{err1}. The mirror surface height error of the SRM is given in Eq.~\eqref{err2}. The horizontal axis is the level of degeneracy, measured by two quantities: the g-factor of the SRC and the one-way Guoy phase in the SRC. The horizontal position of the left-most point of the curve corresponds to the degeneracy level of the SRC in the baseline Advanced-LIGO design.
\label{fig:srm}}
\end{center}
\end{figure}

\subsection{Different modes of deformation on the SRM and ITMs}
\label{srm24}

In this section, we consider ACs and the SRC with different modes of deformation in the SRM and ITMs. In Fig.~\ref{fig:srm},  we show the loss of the SNR as a function of SRC degeneracy level, when there are common curvature radius errors of $\Delta R=1$m on the ITMs, and a fourth-order polynomial deformation on the SRM with the form
\bea\label{err2}
\Delta z(x,y)&=&-4.65\nm+17.94\nm\left[\left(\frac{x}{6\cm}\right)^2+\left(\frac{y}{6\cm}\right)^2\right] \nonumber\\
&&-8.64\nm\left[\left(\frac{x}{6\cm}\right)^2+\left(\frac{y}{6\cm}\right)^2\right]^2 \,.
\eea
This $\Delta z$ is plotted in Fig.~\ref{fig:figerr}.

Comparing Fig.~\ref{fig:srm} with Fig.~\ref{fig:itm2sig}, we see two more HOM resonant peaks generated by the fourth-order Hermite-Gaussian modes at Guoy phase around 0.8 and 1.4 radian. We see from Fig.~\ref{fig:srm} that, although the optical modes in the AC and the SRC are different, and there is an eigenmode mismatch on the ITMs, the nondegenerate SRC does not simply reject the mismatched part of the signal power, and the loss of the SNR is still very low (away from the HOM resonant peaks). This is because the SRC effectively reflects the HOMs that cause the mode mismatch back into the AC and helps the fundamental mode build-up in the AC, so long as the SRC tuning for the fundamental mode resonance is unchanged. (This tuning depends on how the control system works. Modeling of the control sidebands is being considered in more sophisticated simulations that are under development, e.g., Advanced-LIGO FFT simulations \cite{fft}). More precisely, this argument is valid because the coupling of the carrier to the signal sideband happens at the ETM, outside the SRC; if it happened inside the SRC, the non-matching HOMs would be expelled and thus the signal power in the fundamental mode would be reduced.

\begin{figure}
\begin{center}
\includegraphics[width=0.5\textwidth]{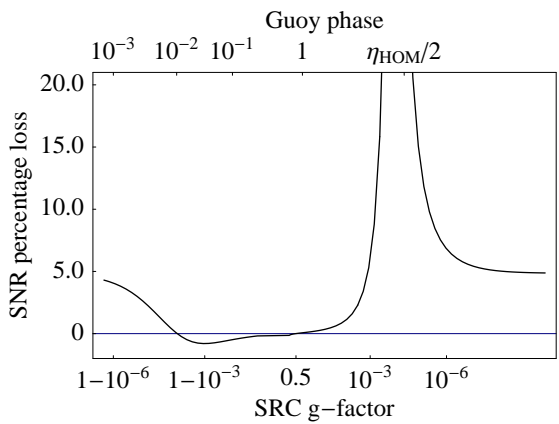}
\caption[Loss of the SNR in narrowband Advanced-LIGO interferometers due to common mirror curvature-radius errors on the ITMs]{Loss of the SNR in narrowband (see Eqs.~\eqref{tuning} and \eqref{trans} for parameters) Advanced-LIGO interferometers due to common mirror curvature radius errors on the ITMs, at the resonant signal sideband frequency $f_0-\lambda/2\pi$, as a function of  SRC degeneracy. The common error of curvature radius is $\Delta R_{\rm ITM}=2$m, i.e., two-fifth the error considered in Figs.~\ref{fig:itm2sig} and \ref{fig:itm2car} or equivalently two-fifth the mirror surface height error given in Eq.~\eqref{err1}. The horizontal axis is the level of degeneracy, measured by two quantities: the g-factor of the SRC and the one-way Guoy phase in the SRC. The horizontal position of the left-most point of the curve corresponds to the degeneracy level of the SRC in the current Advanced-LIGO baseline design.
\label{fig:itmnarrow}}
\end{center}
\end{figure}

\subsection{Curvature radius error on the ITMs: Narrowband configuration}
\label{itm2narrow}

All examples above are simulations for the broadband Advanced-LIGO configuration, i.e., the RSE configuration, in which the signal storage time in the AC is reduced by the SRC. In the narrowband configuration, by contrast, the signal light is truly being ``recycled," and the storage time in the SRC is an order of magnitude longer than in the RSE scheme, which could change our broadband-interferometer results significantly. In Fig.~\ref{fig:itmnarrow}, we show simulation results for the narrowband configuration with differential curvature errors $\Delta R=2$m on the ITMs. The SNR loss for the baseline degenerate SRC is about 5\%, which is very significant. In Ref. \cite{mesa}, the narrowband configuration also implies the most severe constraint on mirror figure errors. The SRC finesse in the narrowband configuration is much higher than in the broadband Advanced-LIGO configuration, so the HOMs being excited inside the SRC are built up to higher power, when the SRC is degenerate. As for other configurations, the loss of the SNR is reduced by orders of magnitude when the SRC is changed from degenerate to nondegenerate.

\subsection{Optimal SRC degeneracy}
\label{optimal}

\begin{figure*}
\begin{center}
\includegraphics[width=1\textwidth]{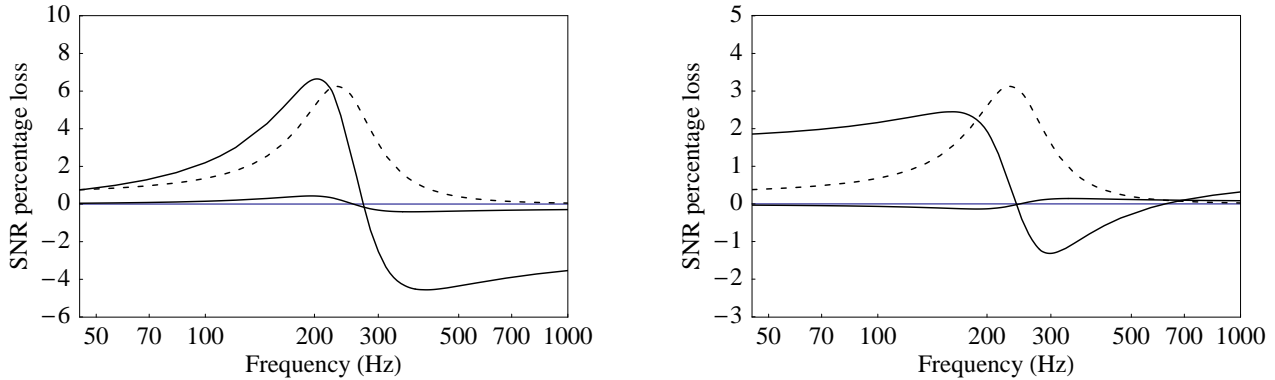}
\caption[Loss of the SNR in an Advanced-LIGO interferometer due to mirror curvature-radius errors on the ITMs, for signal sidebands with various frequencies]{Loss of the SNR in an Advanced-LIGO interferometer due to mirror curvature radius error on the ITMs, for signal sidebands with various frequencies. The mirror curvature errors are the same as those consider in Fig.~\ref{fig:itm2sig}, and we show in the left and right panels the results for common and differential errors. In each case, we consider two designs of the SRC: one degenerate SRC with the Advanced-LIGO baseline design parameters, and one nondegenerate SRC with one-way Guoy phase $\eta=0.38$ (i.e., g-factor $0.86$) prescribed preliminarily in the MMT design \cite{mmt}. In each plot the solid curve with larger variations shows the loss of the SNR when the degenerate SRC is used, while the solid curve with smaller variations shows the loss of the SNR when the nondegenerate SRC is used. The dashed curve, plotted in arbitrary units, shows the amplitude of the output signal sideband light in an ideal Advanced-LIGO interferometer with fixed GW strain. The maximum of the dashed curve is at the frequency $f_0-\lambda_g/2\pi$ with $\lambda_g$ given by Eq.~\eqref{omegas}.
\label{fig:itmspec}}
\end{center}
\end{figure*}

From the examples above, we see that the mode mixing and consequent problems are suppressed by  making the SRC nondegenerate. The optimal Guoy phase in the SRC for Hermite-Gaussian modes is the range 0.2 to 1.3 radians. In the examples above we showed the loss of the SNR only at the most sensitive signal sideband frequency [i.e., $f_0-\lambda_g/2\pi$ given by Eq.~\eqref{omegas}]. In Fig.~\ref{fig:itmspec}, we show the interferometer response of various signal frequencies assuming that the SRC is either degenerate with the current design parameters or nondegenerate with the MMT design suggested by M\"{u}ller \cite{mmt} (the cavity Guoy phase is 0.38 radians, and thus nondegenerate). There is a strong suppression of the SNR loss across the entire Advanced-LIGO sensitive band, and a small shift of the most sensitive frequency (remember however that at low frequency $f<100$Hz, we should also consider the radiation pressure noise).

At first sight, there seems to be a wide optimal range from which to choose the SRC degeneracy, but our freedom is actually quite limited. One obvious constraint is that we need to avoid those values of the Guoy phase that give rise to the HOM resonant peaks. In a realistic interferometer, with many more HOM perturbations, there would be a large number of HOM resonant peaks across the optimal Guoy phase range, so the degeneracy should be chosen carefully, through careful simulations.

However, the worst difficulty posed by the above Guoy phase range is a practical one. There are two obvious ways to achieve the desired g-factor: reduce the beam size in the SRC, or increase the length of the SRC. Unfortunately, to achieve the required low degeneracy, we need either a very small beam waist size near the SRM, or a kilometers-long recycling cavity length. In the next section, we discuss the practical alternative designs of these two types.

\section{Alternative designs}
\label{sec:designs}

There are two obvious ways to reduce the SRC degeneracy: reduce the beam size and/or increase the optical path length. We discuss two ideas in turn.

To reduce the SRC beam size, we could add a lens in the recycling cavity; but to get a Guoy phase between 0.2 and 1.3 radians, the beam must be focused so strongly that the waist size of the beam is of order $3\times10^{-5}$m, and the waist location must be tuned precisely to a few millimeters away from the SRM. For a $1\times10^{-5}$m beam size on the SRM, we will have a $10$GW/m$^2$ power density heating one spot on the mirror, which is too high to be practical. This problem was pointed out qualitatively by Bochner \cite{bochner} in his FFT simulation work.

We can circumvent this problem by introducing multiple steps of beam focusing, i.e., bringing down the beam size step by step with more optical elements, so there is substantial Guoy phase accumulated during each step with small beam size. It is clearly preferable to use reflective optical elements in this scheme. A practical design based on moving the mode-matching telescopes (MMT) into the recycling cavities has been proposed by M\"{u}ller and Wise in Ref.~\cite{mmt}. More specifically, the MMTs outside the recycling cavities used to match the beam size from mm scale to cm scale between the light source and the PRM, and from cm scale to mm scale between the SRM and the photo detector are moved into the cavities to reduce the beam size there. Two mirrors are used in each MMT. The first brings down the beam size from $\sim 6$cm to millimeter scale, and the second tunes the shape of the millimeter-scale beam to achieve the desired degeneracy. Practical parameters for the PRC can be found in Ref.~\cite{mmt}, and the coupled recycling cavities and cavities in the MMT are stable in principle. The MMT introduces more mirrors and cavities into the interferometer, and there are more control problems for the new mirrors. Experimental studies of issues about implementing this MMT design are ongoing.

Another way to reduce the SRC degeneracy is to use a longer SRC, though the length must be on the kilometer scale to achieve the necessary g-factor. A natural idea is to bend the SRC into one of the beam tubes of the ACs, and make it 4km long. This design seems to have the immediate problem of light scattering noise in the crowded arm tubes, but it does have important scientific advantages. The 4km SRC idea has been advocated for a long time, and reducing the SRC degeneracy is just one gain among other advantages. Mizuno \cite{mizuno} suggested using a 4km long SRC to collect power in both signal sidebands and increase the SNR by a factor of 2. Buonanno and Chen \cite{bc1}, in their analysis of beating the SQL with a signal-recycled interferometer, found that the gain in peak sensitivity is vulnerable to optical losses in the short SRC, and km long SRC with fewer light bounces might solve this problem. Moreover, a long SRC introduces a frequency-dependent correlation between the two quadratures of the vacuum field, and might thereby bring interesting changes to the optical noise spectrum of the interferometer. Finally, light scattering in the beam tube may not cause any significant problems according to preliminary estimates, and more investigation of this issue is ongoing.

\section{Conclusions}
\label{sec:conc}

We have set up a simplified model of an Advanced-LIGO interferometer, with mirror deformations in the form of figure errors, and we have simulated the carrier and signal sideband optical fields in this model interferometer by a mode decomposition method. Using our simulations, we have investigated the loss of the SNR when SRCs with various degeneracies are used, and we have found an optimal range of SRC degeneracies that minimizes the loss of the SNR.

For the current degenerate SRC design, we found that the loss of the SNR, due to mode mixing between the fundamental mode and HOMs, is severe, and our results are consistent with the geometric optics estimates made in Ref.~\cite{mesa}. Assuming that the loss of the SNR scales quadratically with the size of the mirror figure error, and requiring the SNR loss due to mirror deformation to be smaller than 1\%, we found a very severe constraint on the mirror deformations. In the broadband Advanced-LIGO design, If the only type of mirror deformation is curvature-radius error, this error must be smaller than 2.5m--7m on the ITMs, which corresponds to peak-to-valley figure errors smaller than 1nm--3nm in the central region of the mirrors. In reality, the mirror deformation is formed by a combination of many spatial modes, and when we consider the next lowest order spatial modes, as depicted in Eq.~\ref{err2} and Fig.~\ref{fig:figerr}, we get a constraint $\sim 4$nm on the SRM. In the narrowband design, the constraint on the ITMs is tightened to $\sim 0.4$nm. Considering the fact that at leading order, the loss of the SNR grows quadratically with the size of the figure errors, and losses due to figure errors with different spatial modes are added linearly, the constraint on mirror figure errors and thermal effects is very difficult to achieve with current technology. Another minor problem we found with the baseline degenerate SRC is that, when there are differential perturbations on the ACs, a huge amount of carrier light in HOMs is coupled to the dark output port and overwhelms the fundamental-mode reference light. To remove this large HOM light at the output requires a reliable output mode cleaner.

We have shown that a nondegenerate SRC could solve this problem, by suppressing the mode mixing and reducing the loss of the SNR by orders of magnitude, in the Advanced-LIGO sensitive band. We have shown also that a nondegenerate SRC does not simply reject mode mismatched light from the AC; it also helps the fundamental mode to build up.

We propose using a nondegenerate SRC in Advanced LIGO with the one-way Guoy phase between 0.2 and 1.3 radians. There are difficulties in achieving this optimal degeneracy in practice, and we have discussed two possible alternative designs for doing so. In the first design, we move the MMT into the recycling cavities and in the second, we use a $4$km long SRC.

A more complete simulation of the optical fields inside an Advanced-LIGO interferometer using FFT propagation methods is under development \cite{fft}. This will effectively include hundreds of HOMs, and will model important physical factors such as thermal effects on the mirrors, and control sideband fields. This new simulation is aimed at mapping out the phase fronts of the light in a very realistic Advanced-LIGO model, to an accuracy of $\sim10^{-6}$. This new simulation, among other goals, will help perfect designs of SRCs with optimal degeneracy.

\acknowledgments
We wish to thank K. Thorne for suggesting this problem and stimulating discussions. We also thank Yanbei Chen, Guido M\"{u}ller, Phil Willems and Hiroaki Yamamote for very useful discussions and interactions. Research led to this paper has been supported by NSF grants PHY-0099568 and PHY-0601459.

\appendix

\section{Abbreviations and symbols}
\label{app:symbol}

There is a large number of abbreviations and symbols used in this chapter. We give a full list of them in this appendix, for easy reference.

\

Abbreviations:

\begin{tabular}{ll}
AC & arm cavity \\
BS & beam-splitter \\
EM &  electromagnetic\\
ETM& end test mass \\
FFT& fast Fourier transform \\
GW& gravitational wave \\
HOM& higher-order mode \\
ITM& input test mass \\
MMT& mode-matching telescope \\
PRC& power-recycling cavity \\
PRM& power-recycling mirror \\
RF& radio frequency \\
RSE& resonant sideband extraction \\
SNR& signal-to-noise ratio \\
SRC& signal-recycling cavity \\
SRM& signal-recycling mirror \\
TCS& thermal compensation system
\end{tabular}

\newpage

\ 

Symbols:

\begin{tabular}{ll}
$\Delta z$ & mirror surface height error \\
$\eta$ & Guoy phase \\
$\lambda$ & carrier light wavelength \\
$\lambda_g$ & resonant signal sideband frequency in the \\
& combined SRC-AC cavity \\
$\omega_0$ & carrier light frequency \\
$\omega_g$ & GW sideband frequency \\
$\phi_\alpha$ & phase shift in cavity ``$\alpha$" \\
$\phi_h$ & phase shift due to GW strain \\
$c$ & speed of light in vacuum \\
$d$ & Michelson arm length difference \\
$f_0$ & carrier light frequency \\
$k$ & carrier light wave number \\
$L$ & Length of arm cavity \\
$l_\alpha$ & common length of recycling cavity ``$\alpha$" \\
$M_\alpha$ & reflection operator of mirror ``$\alpha$" or \\
& of cavity ``alpha" as a compound mirror; \\
& Exceptions: Michelson operators $\MCC$, $\MCD$, \\
& $\MDC$ and $\MDD$ defined in Eq.~\eqref{michop} \\
$P_\alpha$ & one-way propagator in cavity ``$\alpha$" or \\
& propagator of path ``alpha" \\
$r_\alpha$ & amplitude transmissivity of mirror ``$\alpha$" \\
$T_\alpha$ & transmission operator of mirror ``$\alpha$" \\
$t_\alpha$ & amplitude transmissivity of mirror ``$\alpha$" \\
$Z$ & distortion function
\end{tabular}

\ 

\ 

Superscripts:

\begin{tabular}{ll}
B & broadband design \\
C & carrier light \\
N & narrowband design \\
S & signal sideband \\
\end{tabular}

\ 

\ 

\ 

Subscripts:

\begin{tabular}{ll}
ac1 & online arm cavity \\
ac2 & offline arm cavity \\
b & beam-splitter \\
C & common mode in Michelson arms \\
D & differential mode in Michelson arms \\
$\pm d$ & a $\pm d$ trip in one of the Michelson arms \\
e1 & end test mass of online arm cavity \\
e2 & end test mass of offline arm cavity \\
i1 & input test mass of online arm cavity \\
i2 & input test mass of offline arm cavity \\
p & power-recycling mirror \\
prc & power-recycling cavity \\
rt ac1 & round trip in online arm cavity \\
rt ac2 & round trip in offline arm cavity \\
s & signal-recycling mirror \\
src & signal-recycling cavity \\
\end{tabular}

\section{Assumptions and approximations in Advanced-LIGO interferometer model}
\label{app:app}

In this appendix, We list the principal assumptions and approximations adopted in our simplified Advanced-LIGO interferometer model, that are introduced and discussed in various places in the chapter:

(1) We consider only the five lowest-order HOMs and focus mostly on the two lowest-order modes, i.e., modes excited by mirror curvature-radius errors

(2) We assume the sizes of mirrors in the interferometers are large enough for diffraction losses to be negligible.

(3) We assume that the interferometer noise is dominated by the photon shot noise, and ignore radiation-pressure noise.

(4) We assume all degrees of freedom of the interferometer (cavity tuning, optical element alignment, etc.) are fixed to their ideal values as if there were no mirror deformation, except for the tuning of the PRC and the AC. The PRC and AC are tuned to maximize the total carrier light power (sum of power in all optical modes) in the AC, for fixed input light power.

(5) We include in our simulation only the carrier light and one signal sideband, without any sidebands for control purposes.

(6) We choose a microscopic asymmetry between the Michelson arm lengths that sends about 1W of carrier power to the SRM.

(7) We assume that the beam-splitter is perfect.

(8) We adopt the short distance approximation (Sec.~\ref{modal}).

(9) We ignore the difference in phase shifts in the short recycling cavities between the carrier and signal sidebands.

(10) We assume that all mirrors are lossless.

\section{Solving for the optical fields in the interferometer}
\label{app:fields}

In this appendix, we write out the explicit form of the interaction operator $M$ in the Hermite-Gaussian modal space, and write equations in the form of Eq.~\eqref{Eeqn} for all the carrier and signal sideband fields displayed in Fig.~\ref{fig:ifo}. We solve these equations analytically in terms of the input field $\Ein$, the GW strain, and the propagation and interaction operators. The SNR is proportional to the fundamental mode component of the output signal sideband field $\EDout$

The carrier light and the signal light have inputs at different positions (the signal input is effectively at the ETMs), so their solutions are different and have to be treated separately. In the following we will denote with superscripts ``C" for carrier and ``S" for signal. Furthermore, because of their different frequencies, their propagation operators have to be distinguished as well in the AC; however, for the short recycling cavities, since the fractional difference between the carrier and signal frequencies is on the order of $10^{-11}$, the phase difference between carrier and signal light is negligible and we use the same SRC operators. For the same reason, the interaction operators are the same for carrier and signal light, since the perturbation effect has effectively a length scale $\sim10$nm.

For Hermite-Gaussian modes, the matrix element of $M$ in the modal space is given in Ref.~\cite{principle} as
\bea
M_{mn,kl}&=&\langle mn|e^{-ikZ(x,y)}|kl\rangle \nonumber\\
&=&\Big\langle mn\Big|{\rm exp}\Bigl(-ik\sum_{op,qr}|op\rangle Z_{op,qr}\langle qr|\Bigr)\Big|kl\Big\rangle \,, \nonumber\\
\eea
where $Z_{op,qr}$ can be calculated through the following quantities:

(i) $c_{ij}$ is given by the expansion of the distortion function $Z(x,y)$ into Hermite polynomial modes:
\beq
-kZ(x,y)=\sum_{i,j}c_{ij}H_i\left(\frac{\sqrt{2}x}{w(z)}\right)H_j\left(\frac{\sqrt{2}x}{w(z)}\right) \,,
\eeq
in which $w(z)$ is the beam size at the reference surface where the operator is defined;

(ii) $h^{ij}_{st,qr}$ is given by the expansion of Hermite polynomial products:
\beq
H_i(x)H_j(y)H_q(x)H_r(y)=\sum_{st}h^{ij}_{st,qr}H_s(x)H_t(y) \,.
\eeq

(iii) $T^{ij}_{op,qr}$ is given by the integral
\beq
T^{ij}_{op,qr}=\langle op|H_i(x)H_j(y)|qr\rangle|_{z=0} \,,
\eeq
which is given analytically when the mirror size is infinite as
\beq
T^{ij}_{op,qr}=\frac{h^{ij}_{st,qr}}{2}\sqrt{\frac{2^oo!2^pp!}{2^qq!2^rr!}} \,.
\eeq
When the mirror size is finite, $T^{ij}_{op,qr}$ has to be calculated numerically.

(iv) $Z_{op,qr}$ is then given by:
\beq
Z_{op,qr}=-\frac{2}{k}\sum_{i,j}c_{ij}T^{ij}_{op,qr} \,.
\eeq
Therefore, given $Z(x,y)$, which embodies information about the mirror distortion, and the beam size at the optical element $w(z)$, we get the representation of the operators $M_{mn,kl}$ in modal space following the steps above.

Now we turn to the optical fields. As described in Sec.~\ref{sec:ifomodel}, we solve for the carrier field as the first step. Using results for LIGO that have been derived in Ref.~\cite{principle} and Ref.~\cite{modalup}, and following the conventions in Sec.~\ref{sec:ifomodel}, we define the following.

(i) The round-trip propagator in the AC is given by
\bea
P^{\rm C}_{\rm rt\,ac1}&=&\ria\rea\Mia\PCaca\Mea\PCaca \,, \nonumber\\
P^{\rm C}_{\rm rt\,ac2}&=&\rib\reb\Mib\PCacb\Meb\PCacb \,.
\eea
again we use subscripts ``1" and ``2" to denote the online and offline arm cavities.

(ii) The reflection operators of the ACs (for reflecting off the AC from the ITM side) are given by
\bea
\MCaca&=&\ria\left(\Mia'-\frac{\tia^2}{\ria^2}T_{\rm i1}\Mia^{\dagger}P^{\rm C}_{\rm rt\,ac1}(I-P^{\rm C}_{\rm rt\,ac1})^{-1}T_{\rm i1}\right) \,, \nonumber\\
\MCacb&=&\rib\left(\Mib'-\frac{\tib^2}{\rib^2}T_{\rm i2}\Mib^{\dagger}P^{\rm C}_{\rm rt\,ac2}(I-P^{\rm C}_{\rm rt\,ac2})^{-1}T_{\rm i2}\right) \,, \nonumber\\
\eea
where $\Mia'$ and $\Mib'$ are reflection operators of the ITMs as seen from the recycling cavity side.

(iii) The Michelson cavity operators are given by
\bea\label{michop}
\MCC&=&\tbs^2\PCd\MCaca\PCd+\rbs^2\PCmd\MCacb \PCmd \,, \nonumber\\
\MCD&=&\tbs\rbs (\PCd\MCaca \PCd+\PCmd\MCacb \PCmd) \,,
\eea
where $\PCd$ and $\PCmd$ are propagators through the differential length $d$ of the Michelson arms [Eq.~\eqref{dif}]. By combined with $\PCprc$ and $\PCsrc$, i.e., the propagators through the common lengths of the PRC ($l_{\rm prc}=8.34$m) and SRC ($l_{\rm src}=8.327$m), we get the propagators from the PRM to the PRM and the SRM (i.e., the ``common" and ``differential" mode propagators): $\PCprc\MCC\PCprc$ and $\PCsrc\MCD\PCsrc$.

The coupled equations for the carrier fields are then
\bea\label{ECeqn}
\ECs&=&\tbs\PCprc \PCd\ECrea-\rbs\PCprc \PCmd\ECreb \,, \nonumber\\
\ECa&=&\rbs\PCsrc \PCd\ECrea+\tbs\PCsrc \PCmd\ECreb \,, \nonumber\\
\ECsr&=&-\rp\Mp\ECs+\tp T_{\rm p}\Ein \,, \nonumber\\
\ECar&=&-\rs\Ms\ECa \,, \nonumber\\
\ECina&=&\tbs\PCprc \PCd\ECsr+\rbs\PCsrc \PCd\ECar \,, \nonumber\\
\ECinb&=&-\rbs\PCprc \PCmd\ECsr+\tbs\PCsrc \PCmd\ECar \,, \nonumber\\
\ECrea&=&\MCaca\ECina \,, \nonumber\\
\ECreb&=&\MCacb\ECinb \,, \nonumber\\
\eea
and the circulating fields inside the AC are given in terms of $\ECina$ and $\ECinb$ by
\bea
\ECcira&=&\tia\PCaca\left(I-P^{\rm C}_{\rm rt\,ac1}\right)^{-1}T_{\rm i1}\ECina \,, \nonumber\\
\ECcirb&=&\tib\PCacb\left(I-P^{\rm C}_{\rm rt\,ac2}\right)^{-1}T_{\rm i2}\ECinb \,.
\eea

Solving Eq.~\eqref{ECeqn}, we have
\bea\label{ECsol}
\ECsr&=&\tp\Big(I+\rp\Mp\PCprc\MCC\PCprc-\rp\rs\Mp\PCprc\MCD\PCsrc \nonumber\\
&&\times\left(I+\rs\Ms\PCsrc\MCC\PCsrc\right)^{-1}\Ms\PCsrc\MCD\PCprc\Big)^{-1}T_{\rm p}\Ein \,, \nonumber\\
\ECar&=&-\left(I+\rs\Ms\PCsrc\MCC\PCsrc\right)^{-1}\rs\Ms\PCsrc\MCD\PCprc\ECsr \,. \nonumber\\
\eea
All other carrier fields can be easily calculated from $\ECsr$ and $\ECar$.

Assuming a monochromatic GW wave passing through the interferometer and shaking the ETMs differentially with strain $h_0\cos\omega_g t$, the carrier light that is incident on the ETMs (i.e., $\ECcira$ and $\ECcirb$) is coupled to the motion of the ETMs and generates signal sidebands at frequencies $\omega\pm\omega_g$ with the EM fields given by
\bea
\EDsiga&=&i\phi_h\rea\Mea\ECcira \,, \nonumber\\
\EDsigb&=&-i\phi_h\reb\Meb\ECcirb \,,
\eea
where $\phi_h=2kh_0L$ is the phase shift due to the GW strain. These fields are the input for the signal-light field in the interferometer.

For the signal sideband field, we define the round-trip propagator of the AC:
\bea
P^{\rm S}_{\rm rt\,ac1}&=&\ria\rea\Mea\PCaca\Mia\PCaca \,, \nonumber\\
P^{\rm S}_{\rm rt\,ac2}&=&\rib\reb\Meb\PCacb\Mib\PCacb \,,
\eea
and we define for the signal sideband the reflection operators of the ACs $\MDaca$ and $\MDacb$, and Michelson cavity operators: $\MDC$ and $\MDD$, in the same way as their carrier-light counterparts were defined in Eq.~\eqref{michop}, but using signal sideband propagators.

The coupled equations for the signal fields are similar to Eq.~\eqref{ECeqn}; we only need to change all quantities for the carrier fields in Eq.~\eqref{ECeqn} to their counterparts for the signal fields, and change the positions of the input fields in three of the equations:
\bea\label{EDeqn}
\EDsr&=&-\rp\Mp\EDs \,, \nonumber\\
\EDrea&=&\MDaca\EDina+\tia T_{\rm i1}\PDaca\left(I-P^{\rm S}_{\rm rt\,ac1}\right)^{-1}\EDsiga \,, \nonumber\\
\EDreb&=&\MDacb\EDinb+\tib T_{\rm i2}\PDacb\left(I-P^{\rm S}_{\rm rt\,ac2}\right)^{-1}\EDsigb \,. \nonumber\\
\eea

Solving Eq.~\eqref{EDeqn} for $\EDa$ we obtain the output signal field:
\bea
\EDout&=&\ts T_{\rm s}\EDa=\ts T_{\rm s}\left(I+\rs\PDsrc\MDC\PDsrc\Ms\right)^{-1} \nonumber\\
&&\times\left(-\rp\PDsrc\MDD\PDprc\Mp\EDs+\EDsigaa\right) \,,
\eea
where
\bea
\EDs&=&\Big(I+\rp\PDprc\MDC\PDprc\Mp-\rp\rs\PDprc\MDD\PDsrc\Ms \nonumber\\
&&\times\left(I+\rs\PDsrc\MDC\PDsrc\Ms\right)^{-1}\PDsrc\MDD\PDprc\Mp\Big)^{-1} \nonumber\\
&&\times\Big(\EDsigss-\rs\PDprc\MDD\PDsrc\Ms \nonumber\\
&&\qquad\times\left(I+\rs\PDsrc\MDC\PDsrc\Ms\right)^{-1}\EDsigaa\Big) \,,\nonumber\\
\EDsigss&=&\tia\tbs\PDprc\PDd T_{\rm i1}\PDaca\left(I-P^{\rm S}_{\rm rt\,ac1}\right)^{-1}\EDsiga \nonumber\\
&&-\tib\rbs\PDprc\PDmd T_{\rm i2}\PDacb\left(I-P^{\rm S}_{\rm rt\,ac2}\right)^{-1}\EDsigb \,,\nonumber\\
\EDsigaa&=&\tia\rbs\PDsrc\PDd T_{\rm i1}\PDaca\left(I-P^{\rm S}_{\rm rt\,ac1}\right)^{-1}\EDsiga \nonumber\\
&&+\tib\tbs\PDsrc\PDmd T_{\rm i2}\PDacb\left(I-P^{\rm S}_{\rm rt\,ac2}\right)^{-1}\EDsigb \,.\nonumber\\
\eea

According to Eq.~\eqref{snrtosig}, the SNR is proportional to the amplitude of the output signal sideband field $\EDout$ in the fundamental optical mode. The fractional loss of the SNR is then the fractional loss of the fundamental mode amplitude in $\EDout$ when mirror deformations are introduced through the mirror reflection and transmission operators $M_\alpha$ and $T_\alpha$.

\end{document}